\documentclass[a4paper,12pt]{article}
\usepackage{exscale}
\usepackage{amssymb}
\usepackage{epsfig}

\newcommand{\ssection}[1]{\setcounter{equation}{0}\setcounter{footnote}{0}%
\section{#1}}

\newcommand{\equ}[2]{\begin{equation}\label{e#1}#2\end{equation}}
\newcommand{\equa}[2]{\begin{eqnarray}\label{e#1}#2\end{eqnarray}}
\newcommand{\abl}[2]{\frac{\delta #1}{\delta #2}}
\newcommand{\D}{\displaystyle}
\newcommand{\mr}{\mathrm}

\newcommand{\tb}[1]{\textbf{#1}}
\renewcommand{\(}{\left(}
\renewcommand{\)}{\right)}
\renewcommand{\[}{\left[}
\renewcommand{\]}{\right]}

\renewcommand{\~}{\tilde}
\renewcommand{\^}{\hat}
\renewcommand{\phi}{\varphi}
\renewcommand{\theta}{\vartheta}
\newcommand{\lamb}{\bar\lambda_k}
\newcommand{\lam}{\lambda_k}
\newcommand{\del}{\bigtriangleup}
\newcommand{\Z}{Z_{N k}}
\newcommand{\gb}[1][\mu\nu]{\bar{g}_{#1}}
\newcommand{\hb}[1][\mu\nu]{\bar{h}_{#1}}
\newcommand{\Gb}{\bar{G}}
\newcommand{\vint}[1][\bar g]{\int\!\! d^{\;\!4} x\, \sqrt{#1}\:}

\begin{document}
\title{Gauge Dependence of the Effective Average Action in Einstein Gravity}
\vspace{-1ex}
\author{\large{\bf{Sven~Falkenberg}}\\
        \emph{Universit\"at~Leipzig, Institut~f\"ur~Theoretische~Physik}\\
        \emph{Augustusplatz~10, D-04109~Leipzig, Germany}\\
        \emph{e-mail: Sven.Falkenberg@itp.uni-leipzig.de}\\[2ex]
        \large{\bf{Sergei~D.~Odintsov}}\\
        \emph{Dept.~de~Fisica, Universidad~del~Valle}\\
        \emph{A.A.~25360~Cali,~Colombia}\\
        \emph{and}\\
        \emph{Dept.~of~Physics, Hiroshima~University}\\
        \emph{Higashi-Hiroshima,~Japan}\\
        \emph{(on leave from State Tomsk Ped. Univ., 634041 Tomsk, Russia)}\\
        \emph{e-mail: odintsov@quantum.univalle.edu.co}\\
        \emph{        sergei@ecm.ub.es}}

\date{}
\maketitle
\thispagestyle{empty}
\vfill
\begin{abstract}
\noindent
We study the gauge dependence of the effective average action $\Gamma_k$ and
Newtonian gravitational constant using the RG equation for $\Gamma_k$. Then we
truncate the space of action functionals to get a solution of this equation.
We solve the truncated evolution equation for the Einstein gravity in the
De~Sitter background for a general gauge parameter $\alpha$ and obtain a
system of equations for the cosmological and the Newtonian constants. Analysing
the running of the gravitational constant we find that the Newtonian
constant depends strongly on the gauge parameter. This leads to the appearance
of antiscreening and screening behaviour of the quantum gravity. The resolution
of the gauge dependence problem is suggested. For physical gauges like the 
Landau-De~Witt gauge the Newtonian constant shows an antiscreening.
\end{abstract}
\newpage

\ssection{Introduction}
Despite many efforts during the last decades a consistent theory of quantum gravity
(QG) is still lacking. The standard Einstein gravity is known to be
non-renormalizable. But also renormalizable models like $R^2$-gravity (see
\cite{c} for a review) have other serious problems instead of this (e.~g.
presumably non-unitarity of the S-matrix).

In such circumstances it is quite reasonable to consider some of the known
pre-QG theories as effective models which appear from fundamental theories
and which are valid below a definite scale only (e.~g. Planck mass). In such a
spirit one can introduce models to describe QG in the far infrared (at large
distances) \cite{h, i, j}, new models of four-dimensional QG like dilatonic QG
\cite{j} and so on. For example, taking account of IR effects in QG with
cosmological constant an effective model for the conformal factor \cite{h} has
been constructed. Due to strong renormalization effects in the infrared this
model may provide the solution of the cosmological constant problem. On the
other hand, the type of theories to describe the effective IR dynamics in QG
maybe extended to quantum $R^2$-gravity \cite{m} and to quantum supergravity
\cite{n}. Due to the fact that the effective conformal factor model \cite{h}
maybe formulated as renormalizable theory one can calculate the (perturbative)
running of the Newtonian coupling constant in such a theory \cite{k}. Note that
the running of cosmological and Newtonian coupling constants also can be
discussed in other renormalizable theories (GUTs in curved spacetime,
$R^2$-gravity \cite{l}) where results of one-loop calculations of the
corresponding $\beta$-functions are available. Of course, all such
considerations are closely related with the perturbative approach to the
effective action in QG.

However, if one still insists in using the non-renormalizable Einstein theory
as low-energy effective theory for QG other methods should be employed. In
particular, the use of non-perturbative RG in Wilson's effective action approach
\cite{o} maybe quite useful. In such an approach the non-renormalizability of the
theory is no problem since the RG equation governing the evolution of the
effective action from a scale $\Lambda$, where the theory is well defined, to a
smaller scales $k < \Lambda$ can be constructed. Such a finite RG equation (or
evolution equation) maybe consistently formulated in the frame of the effective
average action \cite{d, e, f, p} (for recent discussions on related
non-perturbative RG approaches see \cite{q} and references therein).

In a recent work \cite{a} the scale-dependent gravitational average effective
action $\Gamma_k[g_{\mu\nu}]$ has been discussed in the background field
formalism. The corresponding exact and truncated evolution equations
have been derived, and a special solution of the truncated equation has been
given. The scale dependence of the cosmological and Newtonian constants in
Einstein gravity has been determined (for earlier discussions of the average
effective potential in QG see \cite{r}). The whole study has been done using
the background harmonic gauge (Fock-De~Donder~gauge). It has been found that
the Newtonian coupling constant increases at large distances.

On the same time it is quite well-known that the standard effective action in
the background field method depends on the choice of the gauge fixing
condition (for a review, see \cite{c}). This gauge dependence is controlled by
the corresponding background Ward identities (Slavnov-Taylor identities). Moreover
it is usually assumed in quantum field theory that the effective action (if
non-perturbatively calculated) does not depend on the gauge condition at its
stationary point. Of course, in practice where only one-loop results are easily
available it often happens that the effective action even at stationary points
depends on the gauge fixing. This could lead to an apparent (or intermediate)
gauge dependence also of some physical quantities.

Similarly, in the study of effective average action with the background field
method its gauge dependence is obvious. Such a gauge dependence is controlled
by the background Ward identities again \cite{u}. However it means that results
obtained by investigation of the evolution equation (especially in the truncated
form) necessarily should depend on the gauge choice. Hence it is interesting to
investigate this gauge dependence for a scale-dependent gravitational
action $\Gamma_k[g_{\mu\nu}]$ and for the associated evolution equation.

In the present paper we discuss this question of gauge dependence for Einstein
gravity with cosmological constant in a class of gauges depending on some
parameter $\alpha$ where $\alpha=1$ corresponds to the harmonic gauge.

The paper is written as follows. In the next section, following  \cite{a}, we
briefly review the scale-dependent gravitational action and derive the exact and
truncated RG equations. Section~3 is devoted to the calculation of the
one-loop partition function for Einstein QG on De~Sitter background
(we work in the $\alpha$-parametric class of gauges). This one-loop partition
function actually defines the truncated evolution equation. In section~4 the
evolution equations for the cosmological and gravitational constants are
explicitly obtained with account of the gauge dependence. The running of the
Newtonian coupling constant is analysed in section~5. It is shown that for some
region of the gauge parameter Einstein gravity shows anti\-screening
($\alpha > -2,73$) while for other values of $\alpha$ it shows a screening
effect. The resolution of this unphysical gauge-depending situation is presented
in the final section.

\ssection{Exact and Truncated Renormalization Group Equation}
We start with a short review of the average effective action formalism in
QG; mainly we follow the exposition given in \cite{a}.

First we introduce a scale-dependent modification of the generating functional
for connected Green's functions of QG in four dimensions
\newline
\parbox{13.9cm}{\begin{eqnarray*}
              \lefteqn{\mr{exp}\{W_k\!\[\zeta^{\mu \nu}, \sigma^\mu, \bar\sigma_\mu;
              \beta^{\mu \nu},\tau_\mu; \gb\]\} = }\hspace{3.5cm}    \\
         & &  \int\!\!{\cal D} h_{\mu \nu}\,{\cal D} C^\mu\, {\cal D} \bar{C}_\mu
              \,\mr{exp} \{- S[\gb[]+h] - S_\mr{gf}[h; \gb[]]        \\
         & & -S_\mr{gh}[h, C, \bar{C}; \gb[]] - \Delta_k S[h, C, \bar{C}; \gb[]]
             -S_\mr{source}\}                                         \quad.
              \end{eqnarray*}}\hfill
\parbox{7mm}{\equa{2.1}{}}\newline
Here we decomposed the integration variable $\gamma_{\mu\nu}(x)$ being the
gravitational field into a fixed background metric $\gb$ and into $h_{\mu\nu}$,
the small fluctuations around this background,
\equ{2.2}{\gamma_{\mu\nu}(x) = \gb (x) + h_{\mu\nu} (x) \quad.}
This background field method (for an introduction, see \cite{c}) permits to
replace the integration over $\gamma$ by an integration
over $h$. $C^\mu$ and $\bar{C}_\mu$ are the ghost and anti-ghost fields for
the gravitational field, respectively.

The first term of the action, $S[\gb[]+h]$, denotes the classical part
being invariant under general coordinate transformations $\delta \gamma$ 
generated by the Lie derivative with respect to an arbitrary vector field. 
At the moment we assume that $S$ is positive definite, since the functional 
integral makes sense only if it converges. However, as has been argued 
in \cite{a} this assumption is not necessary for the validity of the evolution
equation which we want to derive (also the use of Lorentzian metric changes this
equation only in an obvious manner).

The gauge fixing term $S_\mr{gf}$ is given by
\equ{2.3}{S_\mr{gf}[h; \gb[]] = \frac{1}{2 \alpha} \vint \gb[]^{\mu \nu} F_\mu
          F_\nu}
with
$$F_\mu = \sqrt{2} \kappa\,{\cal F}^{\alpha\beta}_{\mu}[\gb[]]\, h_{\alpha\beta} =
          \sqrt{2} \kappa \( \delta^\beta_\mu\, \gb[]^{\alpha\gamma} \bar D_\gamma
        - \frac{1}{2}\, \gb[]^{\alpha\beta} \bar D_\mu \)h_{\alpha\beta} \quad. $$
We introduced here the constant $\kappa = (32 \pi \Gb)^{-\frac{1}{2}}$ where
$\Gb$ is the bare Newtonian constant. The derivatives $\bar D_\mu$ are covariant
with respect to the background metric $\gb$.

By $S_\mr{gh}$ we denote the ghost action which can be written for this choice
of gauge fixing with the help of the Faddeev-Popov operator
$${\cal M}[\gamma, \gb[]]^\mu_{\;\;\nu} = \gb[]^{\mu\rho} \gb[]^{\sigma\lambda}
   \bar D_\lambda (\gamma_{\rho\nu} D_\sigma + \gamma_{\sigma\nu} D_\rho) -
   \gb[]^{\rho\sigma} \gb[]^{\mu\lambda} \bar D_\lambda\, \gamma_{\sigma\nu}
   D_\rho$$
as
\equ{2.4}{S_\mr{gh}[h, C, \bar{C}; \gb[]] = -\sqrt{2} \vint \bar C_\mu {\cal M}
         [\gb[]+h, \gb[]]^\mu_{\;\;\nu} C^\nu   \quad.}
That part of the action introduced so far, namely
$S + S_\mr{gf} + S_\mr{gf}$, is invariant under BRST transformations which will
not be specified here (see \cite{a}).

In addition we introduced an IR cutoff for the gravitational field and for the
ghosts:
\newline
\parbox{13.9cm}{\begin{eqnarray*}
                \Delta_k S[h, C, \bar{C}; \gb[]] &=& \frac{1}{2} \kappa^2 \vint
                h_{\mu\nu} R_k^\mr{grav}[\gb[]]^{\mu\nu\rho\sigma} h_{\rho\sigma} \\
          & & + \sqrt{2} \vint \bar C_\mu R_k^\mr{gh}[\gb[]]\, C^\mu \quad.
                \end{eqnarray*}}\hfill
\parbox{7mm}{\equa{2.5}{}}\newline
$R_k^\mr{grav}[\gb[]]$ and $R_k^\mr{gh}[\gb[]]$ are the cutoff operators being
defined below. They discriminate between low- and high-momentum modes. Whereas
eigenmodes of $-\bar D^2$ with eigenvalues $p^2 \gg k^2$ are integrated out in
(\ref{e2.1}) without suppression, modes with eigenvalues $p^2 \ll k^2$ are
damped by a kind of momentum dependent mass term. This ensures the
renormalizibility of the theory, which is an effective one at scale $k$. Both
cutoff operators have the structure
\equ{2.6}{R_k[\gb[]] = {\cal Z}_k k^2 R^{(0)}(-\bar D^2/ k^2) \quad.}
The dimensionless function $R^{(0)}$ will be chosen as (see \cite{a, d, e, f})
\equ{2.7}{R^{(0)}(u) = \frac{u}{e^u - 1} \quad,}
it interpolates smoothly between $\lim\limits_{u \to 0} R^{(0)}(u) = 1$ and
$\lim\limits_{u\to\infty} R^{(0)}(u) = 0$. The factors ${\cal Z}_k$ are different
for the graviton and for the ghosts. For the ghosts ${\cal Z}_k \equiv
Z_k^\mr{gh}$ is a pure number but for the graviton ${\cal Z}_k \equiv
{\cal Z}_k^\mr{grav}$ is a tensor constructed from the background metric. In the
simplest case it is $({\cal Z}_k^\mr{grav})^{\mu\nu\rho\sigma} = \gb[]^{\mu\rho}
\gb[]^{\nu\sigma} Z_k^\mr{grav}$.

The last term in (\ref{e2.1}) is the source term
\newline
\parbox{13.9cm}{\begin{eqnarray*}
  S_\mr{source} &=& - \vint \{\zeta^{\mu\nu} h_{\mu\nu} + \bar \sigma_\mu
                    C^\mu + \sigma^\mu \bar C_\mu                            \\
                & & \hspace{2cm} + \beta^{\mu\nu} {\cal L}_C(\gb + h_{\mu\nu})
                    + \tau_\mu C^\nu \partial_\nu C^\mu \}
                \end{eqnarray*}}\hfill
\parbox{7mm}{\equa{2.8}{}}\newline
where ${\cal L}_C$ denotes the Lie derivative with respect to the ghost field.
Obviously, the sources $\beta^{\mu\nu}$ and $\tau_\mu$ couple to the BRST
variations of $h_{\mu\nu}$ and $C^\mu$.

Now, to obtain the effective average action $\Gamma_k$, we  perform a Legendre
transformation of $W_k$. Therefore, let us first introduce the $\emph{k}$
\emph{-dependent} classical fields (also called mean fields)
\equ{2.9}{\hb     = \frac{1}{\sqrt{\gb[]}} \abl{W_k}{\zeta^{\mu\nu}} \quad,\quad
          \xi^\mu = \frac{1}{\sqrt{\gb[]}} \abl{W_k}{\bar\sigma_\mu} \quad,\quad
      \bar\xi_\mu = \frac{1}{\sqrt{\gb[]}} \abl{W_k}{\sigma^\mu} \quad.}
Then the effective average action is obtained as
\newline
\parbox{13.9cm}{\begin{eqnarray*}
       \Gamma_k [g, \gb[], \xi, \bar\xi; \beta, \tau]\!\!
   &=& \!\!\!\!\vint\, \{\zeta^{\mu\nu} \hb + \bar\sigma_\mu \xi^\mu + \sigma^\mu
       \bar\xi_\mu \} - W_k[\zeta, \sigma, \bar\sigma; \beta,\tau; \gb[]]        \\
   & & - \Delta_k S[g- \gb[], \xi, \bar\xi; \gb[]] \quad.
                \end{eqnarray*}}\hfill
\parbox{7mm}{\equa{2.10}{}}\newline
The first two terms correspond to the usual Legendre transform. By definition
the effective average action is the difference between the usual Legendre
transform and the cutoff action $\Delta_k S$ with the classical fields inserted.
For $k \to 0$ the cutoff $\Delta_k S$ disappears and we get back the usual
Legendre transform. In analogy to $\gamma_{\mu\nu}$ we introduced the metric
\equ{2.12}{g_{\mu\nu}(x) = \gb(x) + \hb(x)}
and we considered $\Gamma_k$ as functional of $g$ instead of $\hb[]$. The
effective average action $\Gamma_k$ depends on the mean fields as well as
parametrically on the sources $\beta$ and $\tau$ and on the background metric
$\gb[]$. It should be noted that this functional by virtue of the background
field technique is invariant under general coordinate transformations.
In this effective average action excitations with momenta larger than $k$ are
integrated out, while those with momenta smaller than $k$ are suppressed by
$\Delta_k S$ such that $\Gamma_k$ describes an effective theory at scale $k$.
In this sense the flow of $\Gamma_k$ with respect to $k$ interpolates between
an action containing the classical action $\Gamma_{k\to\infty} \equiv S +
S_\mr{gf} + S_\mr{gh} - \vint\{\beta^{\mu\nu} {\cal L}_\xi g_{\mu\nu} + \tau_\mu 
\xi^\nu \partial_\nu \xi^\mu\} \equiv \~S$ and the common effective action 
$\Gamma_{k\to 0} \equiv  \Gamma$.

The trajectory in the space of all action functionals can be obtained as the
solution of renormalization group equations. $S$ enters via the initial
condition for the trajectory at some UV cutoff scale $\Lambda$:
$\Gamma_\Lambda = \~S $. While for renormalizable theories $\Lambda$ may be send
to infinity at the end, for QG it is a fixed, finite scale.

Having now introduced the effective average action we derive the evolution
equation for $\Gamma_k$ according to the procedure introduced in
\cite{d, e, f, p, a}. We take the derivative of the eq.~(\ref{e2.1}) with
respect to the renormalization group "time" $t=\ln k$ and, after Legendre
transformation, express the RHS in terms of $\Gamma_k$ using the relation
between connected and 1PI two-point functions. Thus we obtain the following
\emph{exact} evolution equation
\newline
\parbox{13.9cm}{\begin{eqnarray*}
              \partial_t \Gamma_k
          &=& \frac{1}{2} \mr{Tr}\!\[\(\Gamma_k^{(2)}  + \^R_k\)^{-1}_{\hb[]\hb[]}
              \(\partial_t \^R_k\)_{\hb[]\hb[]} \]                  \\
          &-& \frac{1}{2} \mr{Tr}\!\[\bigg\{\(\Gamma_k^{(2)}  +
              \^R_k\)^{-1}_{\bar\xi\xi}  -  \(\Gamma_k^{(2)}  +
              \^R_k\)^{-1}_{\xi\bar\xi}\bigg\}
              \(\partial_t \^R_k\)_{\bar\xi\xi} \]                  \quad,
                \end{eqnarray*}}\hfill
\parbox{7mm}{\equa{2.15}{}}\newline
where $\Gamma_k^{(2)}$ denotes the second functional derivative of $\Gamma_k$,
e.g.
\equ{2.16}{\( \(\Gamma_k^{(2)}\)_{\xi\bar\xi} \)_\mu^{\quad\nu}(x, y) =
           \frac{1}{\sqrt{\gb[](x)}}\, \frac{\delta        }{\delta    \xi^\mu(x)}\,
         \(\frac{1}{\sqrt{\gb[](y)}}\, \frac{\delta\Gamma_k}{\delta\bar\xi_\nu(y)}
           \)\quad.}
$\^R_k$ is a matrix in the field space whose non-zero entries are
\newline
\parbox{13.9cm}{\begin{eqnarray*}
                \(\^R_k\)_{\hb[]\hb[] }^{\mu\nu\rho\sigma} &=& \kappa^2
                 (R_k^\mr{grav}[\gb[]])^{\mu\nu\rho\sigma}            \\
                \(\^R_k\)_{\bar\xi\xi}                     &=& \sqrt{2}
                \,R_k^\mr{gh}[\gb[]] = - \(\^R_k\)_{\xi\bar\xi}     \quad.
                \end{eqnarray*}}\hfill
\parbox{7mm}{\equa{2.14}{}}\newline
The traces in (\ref{e2.15}) have to be interpreted as $\vint$; they are
convergent, in the UV- and in the IR-region, for all cutoffs which are similar
to (\ref{e2.7}). The main contributions come from the region of momenta around
$k$, because of the appearance of the factor $\partial_t \^R_k$; large momenta
are exponentially suppressed.

Now the problem is to solve eq.~(\ref{e2.15}). One way to find at least
approximate but still nonperturbative solutions is to truncate the space of
action functionals. This means that we work on a finite-dimensional subspace
parametrised by only a few  couplings. Let us in a first step try to
neglect the evolution of ghosts. This leads us to the following ansatz
\newline
\parbox{13.9cm}{\begin{eqnarray*}
                \Gamma_k [g, \gb[], \xi, \bar\xi; \beta, \tau] &=& \bar\Gamma_k[g]
             +\^\Gamma_k [g, \gb[]] + S_\mr{gf}[g-\gb[]; \gb[]]
             +S_\mr{gh}[g-\gb[], \xi, \bar\xi; \gb[]]                       \\
          & &-  \vint \{\beta^{\mu\nu} {\cal L}_\xi\, g_{\mu\nu} + \tau_\mu
                \xi^\nu\, \partial_\nu \xi^\mu\}                            \quad.
                \end{eqnarray*}}\hfill
\parbox{7mm}{\equa{2.17}{}}\newline
The classical ghost and gauge fixing terms have the same functional form as for
the bare action. Also the couplings to the BRST variations do not change. The
other terms depend on $g$ and $\gb[]$. The first term
\equ{2.18}{\bar\Gamma_k[g] \equiv \Gamma_k[g, g, 0, 0; 0, 0]}
is the most interesting one, while $\^\Gamma_k$ contains only the deviations for
$\gb[] \neq g$ such that $\^\Gamma_k[g, g]=0$. This leads to the interpretation
of $\^\Gamma_k$ as describing the quantum corrections to the gauge fixing term
$S_\mr{gf}$ which also vanishes for $\gb[] = g$.

It is easy to show that for an UV cutoff scale $\Lambda \rightarrow \infty$ the
action $\bar\Gamma_k$ fulfils the initial condition $\bar\Gamma_\Lambda[g] =
S[g]$ with $S$ being the classical action. If we solve eq.~(\ref{e2.15}) directly
and analyse the initial condition for $\Gamma_k$ we find that
$\^\Gamma_\Lambda = 0$ must be satisfied.

This suggests to set $\^\Gamma_k = 0$ for all $k$ in a second step. From the 
Slavnov-Taylor identity of $\Gamma_k$ it follows (for details see
\cite{a}) that $\^\Gamma_k$ contains higher loop effects such that this kind of
approximation really makes sense.

Inserting ansatz (\ref{e2.17}) into the evolution equation (\ref{e2.15}) we
obtain the \emph{truncated} evolution equation which we wanted to derive
\newline
\parbox{13.9cm}{\begin{eqnarray*}
              \partial_t \Gamma_k[g, \gb[]]
          &=& \frac{1}{2} \mr{Tr}\!\[\( \kappa^{-2} \Gamma_k^{(2)}[g, \gb[]] +
               R_k^\mr{grav}[\gb[]] \)^{-1} \partial_t R_k^\mr{grav}[\gb[]] \]   \\
          & & \:\!\!-     \mr{Tr}\!\[\(- {\cal M}[g, \gb[]] +
               R_k^\mr{gh  }[\gb[]] \)^{-1} \partial_t R_k^\mr{gh  }[\gb[]] \]
              \quad.
              \end{eqnarray*}}\hfill
\parbox{7mm}{\equa{2.19}{}}\newline
Here we introduced
\equ{2.20}{\Gamma_k[g, \gb[]] = \bar\Gamma_k[g] + S_\mr{gf}[g-\gb[]; \gb[]]
           \simeq \Gamma_k[g, \gb[], 0, 0; 0, 0]}
while $\Gamma_k^{(2)}$ is the Hessian of $\Gamma_k[g, \gb[]]$ with respect to
$g_{\mu\nu}$ at a fixed background metric $\gb$. As a result of our ansatz the
LHS of (\ref{e2.19}) does not contain any contributions from the ghosts.

\ssection{One-Loop Quantum Gravity on the De Sitter Background}
In this section we will find the one-loop partition function for Einstein QG.
The classical action is written as
\equ{3.1}{S = \frac{1}{16 \pi \Gb} \vint[g] \[-R(g) + 2 \bar\lambda\] \quad, }
where non-renormalizable Einstein gravity is considered to be valid below a UV
scale $\Lambda$. The truncation
\equ{3.2}{\Gb \to G_k = \frac{\Gb}{\Z}, \quad \bar\lambda \to \lamb, \quad
          \alpha \to \frac{\alpha}{\Z}}
is supposed to be done. In general also $\alpha$ depends on $k$. In this case we
would have three $k$-depending parameters instead of two. This would lead to a
more complicated system of equations for these parameters which could be solved
with numerical methods only. For the pure Yang-Mills theory such a calculation has 
been done \cite{x}. We restrict ourselves to the case that $\alpha$ is a constant.

Substituting (\ref{e3.2}) in (\ref{e2.20}) we easily get \cite{a}
\equ{3.2.1}{\begin{array}{r@{}l}
            \D \Gamma_k[g, \gb[]] = 2\kappa^2 \Z
         &  \D \vint[g] \[-R(g) + 2 \lamb\]                             \\[.8em]
            \D+\,\frac{\kappa^2}{\alpha} \Z
         &  \D \vint \, \gb[]^{\mu\nu} \( {\cal F}_\mu^{\alpha\beta}
             g_{\alpha\beta}\)\({\cal F}_\nu^{\rho\sigma} g_{\rho\sigma}\)
            \quad.
            \end{array}}

For the RHS of the evolution equation we need the second functional derivative of
$\Gamma_k[g, \gb[]]$. In order to calculate it we should specify some curved 
background. A useful choice is the De Sitter space which is maximally symmetric. 
This means that Riemann and Ricci tensors can be written as
\newline
\parbox{13.9cm}{\begin{eqnarray*}
                \bar R_{\mu\nu\rho\sigma}
     &=&        \frac{1}{12} \[ \gb[\mu\rho]  \,\gb[\nu\sigma] -
                                \gb[\mu\sigma]\,\gb[\nu\rho]  \] \bar R  \quad,\\
                \bar R_{\mu\nu}
     &=&        \frac{1}{4}\, \gb \, \bar R                              \quad,
                \end{eqnarray*}}\hfill
\parbox{7mm}{\equa{3.4}{}}\newline
where $\bar R$ is the curvature which works as an external parameter characterising
the space. Using the decomposition
\equ{3.5}{\hb = \^h_{\mu\nu} + \frac{1}{4}\, \gb\, \phi \quad, \quad
          \bar    g^{\mu\nu} \, \^h_{\mu\nu} = 0}
we find
\equ{3.6}{\begin{array}{r@{}l}
          \D\Gamma_{k\,\mr{grav}}^{(2)} = 2 \Z\, \kappa^2\!\! \vint\!
          \left\{    \frac{1}{2}\, \^h_{\mu\nu} \bigg[\! \right.
     &    \D\left. - \bar\square - 2 \lamb + \frac{2}{3} \bar R \bigg]
\^h^{\mu\nu}\right.\\
          \D\left. - \frac{1}{8}\,       \phi      \bigg[\! \right.
     &    \D\left. - \bar\square - 2 \lamb\bigg] \phi  - \[ \bar D^\nu \^h_{\mu\nu}
                   - \frac{1}{4}\, \bar D_\mu \phi\]^2  \right\} ,
          \end{array}}
\equ{3.7}{\Gamma_{k\,\mr{gf}}^{(2)} = 2 \Z\, \kappa^2 \frac{1}{\alpha} \vint\!
          \[ \bar D^\nu \^h_{\mu\nu}  - \frac{1}{4}\, \bar D_\mu \phi\]^2 \quad.}
However, only for the minimal choice $\alpha=1$, the non-minimal terms
in the sum of (\ref{e3.6}) and (\ref{e3.7}) cancel each other. For other choices of
$\alpha$ we are dealing with non-minimal operators. In order to make a heat kernel 
expansion with respect to the curvature we need to work with canonical operators like 
$\del = \square + m^2$ only.

We solve this problem making a further decomposition of the quantum fields.
Like in \cite{b} we write
\equ{3.9}{\^h_{\mu\nu} = \^h_{\mu\nu}^\perp + \(\bar D_\mu \xi_\nu^\perp +
          \bar D_\nu \xi_\mu^\perp\) + \(\bar D_\mu \bar D_\nu - \frac{1}{4} \gb
          \bar \square \) \sigma}
where $D^\mu \xi_\mu^\perp = 0$ and $D^\mu \^h_{\mu\nu}^\perp = 0$. 
It is necessary then to define the differentially constrained operators
($\psi$ is an arbitrary scalar)
\newline
\parbox{13.9cm}{\begin{eqnarray*}
                \del_0                          (X)  \psi
            &=& \(-\bar    \square +             X\) \psi      \quad,\\
                \del_{1  \mu\nu}                (X)  \xi^{\nu \perp}
            &=& \gb\(-\bar \square +             X\) \xi^{\nu \perp} \qquad\mbox{and}\\
                \del_2{}_{ \alpha\beta}^{\mu\nu}(X)  \^h_{\mu\nu}^\perp
            &=& \delta_\alpha^\mu \delta_\beta^\nu
                   \(-\bar \square +             X\) \^h_{\mu\nu}^\perp \quad.
                \end{eqnarray*}}\hfill
\parbox{7mm}{\equa{3.10}{}}\newline
Note that a similar technique has been discussed in \cite{g}.

Now the one-loop partition function in the \emph{gravitational sector} can be easily
calculated \cite{c, b, g}
\equ{3.11}{\begin{array}{r@{}l}
Z_\mr{grav} =& \D\int {\cal D}\^h_{\mu\nu} {\cal D} \phi\, \exp\[-
               \Gamma_{k\,\mr{grav}}^{(2)}-\Gamma_{k\,\mr{gf}}^{(2)}\]  \\[.8em]
            =& \D\[\det \del_1\!\(\frac{1}{4} \bar R\(2\alpha-1\) - 2\alpha\lamb\)
               \]^{ - \frac{1}{2}}
                 \[\det \del_2\!\(\frac{2}{3} \bar R              - 2      \lamb\)
               \]^{ - \frac{1}{2}}                                      \\[.8em]
        \cdot& \D\[\det \del_0\!\(\frac{1}{2} \bar R\( \alpha-1\) - 2\alpha\lamb\)
               \]^{- \frac{1}{2}} \;\!\!\cdot
             \bigg[\det \del_0\!\(                                - 2      \lamb\)
             \bigg]^{-\frac{1}{2}} \quad.
           \end{array}}
Here, in the course of the calculations the decomposition (\ref{e3.9}) is
done and the corresponding Jacobian is taken into account (for more details,
see \cite{b}).

As we see the one-loop partition function is expressed in terms of constrained
operators. However, the transition to constrained operators according to
(\ref{e3.9}) and (\ref{e3.10}) introduces \emph{additional} zero modes. This 
leads to a wrong answer when we expand the determinants of the constrained 
operators in (\ref{e3.11}) as powers of the curvature using the heat kernel 
expansion.  Therefore we will present $Z_\mr{grav}$ in terms of unconstrained 
operators. It could be done using the following relations \cite{b}:
\newline
\parbox{13cm}{\begin{eqnarray*}
              \det \del_V(X) \equiv \det (-\bar\square+X)_V
        &=&   \det \del_1(X) \det \del_0\!\(X-\frac{\bar R}{4}\)           \quad,\\
              \det \del_T(X) \equiv \det (-\bar\square+X)_T
        &=&   \det \del_2(X) \det \del_1\!\(X-\frac{5}{12}\bar R\)               \\
        & &   \!\!\!\cdot    \det \del_0\!\(X-\frac{2}{3} \bar R\)         \quad.
              \end{eqnarray*}}\hfill
\parbox{7mm}{\equa{3.12}{}}\newline
Finally, we obtain
\equ{3.13}{\begin{array}{r@{}l}
Z_\mr{grav} =& \D\[\det \del_T\!\(\frac{2}{3} \bar R             - 2      \lamb\)
               \]^{ - \frac{1}{2}}
                 \[\det \del_V\!\(\frac{1}{4}\(2\alpha-1\)\bar R - 2\alpha\lamb\)
               \]^{ - \frac{1}{2}}                                       \\[.8em]
        \cdot& \D\[\det \del_V\!\(\frac{1}{4} \bar R             - 2      \lamb\)
               \]^{  \frac{1}{2}} \;\:
             \bigg[\det \del_S\!\(                               - 2      \lamb\)
             \bigg]^{-\frac{1}{2}} \quad.
           \end{array}}

In order to calculate the complete partition function we should also take into
account the ghost contribution. The ghost action (\ref{e2.4}) simplifies for
the De Sitter space and for $\gb[] = g$ (this corresponds to a special choice of
the metric field) to the standard form
\equ{3.14}{S_\mr{gh} = -\sqrt{2} \vint \,\bar C_\mu \[-\square -
           \frac{1}{4} R\] C^\mu   \quad.}
The corresponding partition function can be easily found:
\equ{3.15}{Z_\mr{gh} = \det \del_V\!\(-\frac{1}{4} R \)  \quad.}
The complete one-loop partition function for the Einstein QG in the gauge
(\ref{e2.3}) is the product of (\ref{e3.13}) and (\ref{e3.15}) where we set
$\gb[] = g$.

For the truncated evolution equation (\ref{e2.19}) we have to include the cutoff term
$\Delta_k S$ into our calculation. This means that we have to calculate $W_k
\equiv \ln Z_k$ where the path integral of $Z_k$ contains in addition to
$\Gamma_{k\,\mr{grav}}^{(2)}$, $\Gamma_{k\,\mr{gf}}^{(2)}$ and the ghost term
also $\Delta_k S$.

The coefficients ${\cal Z}_k^\mr{grav}$ and $Z_k^\mr{gh}$ have to be chosen
such that the kinetic term and the cutoff term combine to $- D^2 + k^2
R^{(0)}(- D^2/k^2)$ for every degree of freedom. $R^{(0)}$ has to be seen
as a dimensionless function. Because we did not make any decomposition of the
ghosts and neglected all renormalization effects of the ghosts we can set
$Z_k^\mr{gh} \equiv 1$. Through the decomposition of the gravitational field into
four parts the kinetic term of every part obtained an additional factor. These
factors have to be compensated by ${\cal Z}_k^\mr{grav}$ containing projectors
on every part. Also ${\cal Z}_k^\mr{grav}$ is proportional to the renormalization
factor $\Z$.

Then the truncated evolution equation (\ref{e2.19}) reads ($\gb[]=g$)
\equ{3.16}{\begin{array}{r@{}l}
\D\partial_t \Gamma_k[g, g] =
    \D  \frac{1}{2} \mr{Tr}_T 
 &  \D\[{\cal N}\(-\square+\frac{2}{3}        R - 2      \lamb + k^2 R^{(0)} \)^{-1}\]   \\[1em]
    \D+ \frac{1}{2} \mr{Tr}_V 
 &  \D\[{\cal N}\(-\square+\frac{2\alpha-1}{4}R - 2\alpha\lamb + k^2 R^{(0)} \)^{-1}\]   \\[1em]
    \D- \frac{1}{2} \mr{Tr}_V 
 &  \D\[{\cal N}\(-\square+\frac{1}{4}        R - 2      \lamb + k^2 R^{(0)} \)^{-1}\]   \\[1em]
    \D+ \frac{1}{2} \mr{Tr}_S \,
 &  \D\[{\cal N}\,\bigg(\!\!\!\:-\!\square             -2\lamb + k^2 R^{(0)}\bigg)^{-1}\]\\[1em]
    \D- \;\:\,      \mr{Tr}_V 
 &  \D\[{\cal N}_0\!\!\!\;\(-\square-\frac{1}{4}  R            + k^2 R^{(0)} \)^{-1}\] 

           \end{array}}
with 
\equ{4.3}{{\cal N}(z)   = \frac{\partial_t\!\[ \Z\, k^2 R^{(0)}(z) \]}{\Z} \quad,
           \quad  {\cal N}_0(z) = \partial_t\!\[ k^2 R^{(0)}(z) \]         \quad.}
Here the variable $z$ replaces $-D^2/k^2$. 

\ssection{Evolution Equations for Newtonian and Cosmological Constant}
In this section we write down the renormalization group equation (\ref{e2.19})
for the action (\ref{e3.2.1}). We differentiate (\ref{e3.2.1}) with respect to
$t$, then we set $\gb = g_{\mu\nu}$. This corresponds to a special choice of a
field theory. It follows that the gauge fixing term $S_\mr{gf}$ vanishes such
that the LHS of the renormalization group equation reads
\equ{4.1}{\partial_t \Gamma_k[g, g] = 2 \kappa^2\!\vint[g] \[-R(g)\,
          \partial_t \Z + 2\,\partial_t (\Z \lamb) \]   \quad.}
Here the dependence on the gauge fixing parameter $\alpha$ disappears.

Now we want to find the RHS of the evolution equation. We expand the operators 
in (\ref{e3.16}) with respect to the curvature $R$ because we are only interested
in terms of order $\vint[g]$ and $\vint[g] R$:
\newline
\parbox{13.9cm}{\begin{eqnarray*}
                \del_i^{-1}\!\(b\,R-2 a\lamb + k^2 R^{(0)} \)
           &=&  \del_i^{-1}\!\(   -2  a\lamb + k^2 R^{(0)} \)    \\
           &-&  \del_i^{-2}\!\(   -2  a\lamb + k^2 R^{(0)} \) b\,R + O\!\(R^2\)
                \end{eqnarray*}}\hfill
\parbox{7mm}{\equa{4.4}{}}\newline
where $a$ and $b$ are free constants. For a more compact notation let us
introduce
\equ{4.5}{\~\del_{i a}(z) = \del_i\!\(-2  a\lamb + k^2 R^{(0)}(z)\) \quad.}
These steps lead then to
\newline
\parbox{13.9cm}{\begin{eqnarray*}
           \partial_t \Gamma_k[g, g]
&=&        \frac{1     }{2}\mr{Tr}_T\!\[{\cal N}  \~\del_{T 1     }^{-1}\] +
           \frac{1     }{2}\mr{Tr}_V\!\[{\cal N}  \~\del_{V \alpha}^{-1}\] -
           \frac{1     }{2}\mr{Tr}_V\!\[{\cal N}  \~\del_{V 1     }^{-1}\]       \\
&+&        \frac{1     }{2}\mr{Tr}_S\!\[{\cal N}  \~\del_{S 1     }^{-1}\] -
           \;\;            \mr{Tr}_V\!\[{\cal N}_0\~\del_{V 0     }^{-1}\]       \\
&-&R\left\{\frac{1     }{3}\mr{Tr}_T\!\[{\cal N}  \~\del_{T 1     }^{-2}\] +
        \frac{2\alpha-1}{8}\mr{Tr}_V\!\[{\cal N}  \~\del_{V \alpha}^{-2}\]\right.\\
& &\left.\;\;\,-\frac{1}{8}\mr{Tr}_V\!\[{\cal N}  \~\del_{V 1     }^{-2}\] +
   \qquad\:\frac{1     }{4}\mr{Tr}_V\!\[{\cal N}_0\~\del_{V 0     }^{-2}\]\right\}+
           O\!\(R^2\)      \quad.
                \end{eqnarray*}}\hfill
\parbox{7mm}{\equa{4.6}{}}
The terms with ${\cal N}_0$ are the contributions of the ghosts.

As a next step we evaluate the traces. We use the heat kernel expansion
which for an arbitrary function of the covariant Laplacian $W(D^2)$ reads
\equ{4.8}{\begin{array}{r@{}l}
          \D\mr{Tr}_j\!\[W(-D^2)\] = \(4\pi\)^{\!-2} \mr{tr}_j(I)
    &     \D\left\{ Q_2[W] \vint[g]                          \right.  \\[0.8em]
   +&     \D\left.\frac{1}{6} Q_1[W] \vint[g] R + O\!\(R^2\) \right\} \quad;
          \end{array}}
by $I$ we denote the unit matrix in the space of fields on which $D^2$ acts.
Therefore $\mr{tr}_j (I)$ simply counts the number of independent degrees
of freedom of the field of sort $j=T, V, S$, namely
$$\mr{tr}_S(I) = 1, \quad \mr{tr}_V(I) = 4, \quad \mr{tr}_T(I) = 9 \quad.$$
The sort $j$ of fields enters (\ref{e4.8}) via $\mr{tr}_j(I)$ only. Therefore we
will drop the index $j$ of $\~\del_{j a}$ after the evaluation of the traces in
the heat kernel expansion.

The functionals $Q_n$ are the Mellin transforms of $W$,
\newline
\parbox{13.9cm}{\begin{eqnarray*}
                Q_0[W] &=& W(0)                                       \\
                Q_n[W] &=& \frac{1}{\Gamma(n)} \int_0^\infty\!\! d z\,
                           z^{n-1} W(z) \quad,\quad (n > 0)     \quad.
              \end{eqnarray*}}\hfill
\parbox{7mm}{\equa{4.9}{}}

Now we have to perform the heat kernel expansion (\ref{e4.8}) in
eq.~(\ref{e4.6}). This leads to a polynomial in $R$ which is the RHS of the
evolution equation. By comparison of coefficients with the LHS of the evolution
equation (\ref{e4.1}) we obtain in order of $\vint[g]$
\equ{4.10}{\partial_t\!\(\Z \lamb\) = \frac{1}{8 \kappa^2}\,\frac{1}{\(4 \pi\)^2}
           \bigg\{ 6\, Q_2\!\[{\cal N}  \~\del_{1   }^{-1}\] +
                   4\, Q_2\!\[{\cal N}  \~\del_\alpha^{-1}\] -
                   8\, Q_2\!\[{\cal N}_0\~\del_{0   }^{-1}\] \bigg\} }
and in order of $\vint[g] R$
\equ{4.11}{
\begin{array}{r@{}l}
           \D\partial_t\Z = - \frac{1}{24 \kappa^2}\,\frac{1}{\(4 \pi\)^2}
\bigg\{&      -\,30\,Q_2\!\[{\cal N}  \~\del_{1   }^{-2}\] \:-  6\(2\alpha-1\)
                     Q_2\!\[{\cal N}  \~\del_\alpha^{-2}\]              \\[0.8em]
       &      -\,12\,Q_2\!\[{\cal N}_0\~\del_{0   }^{-2}\]   +  6
                   \,Q_1\!\[{\cal N}  \~\del_{1   }^{-1}\]              \\[0.6em]
       &  +\;\:\, 4\,Q_1\!\[{\cal N}  \~\del_\alpha^{-1}\] \:-  8
                   \,Q_1\!\[{\cal N}_0\~\del_{0   }^{-1}\] \bigg\}      \quad.
\end{array}}
We introduce the cutoff dependent integrals
\newline
\parbox{13.9cm}{\begin{eqnarray*}
                \Phi_n^p(w)
            &=& \frac{1}{\Gamma(n)} \int_0^\infty\! dz\, z^{n-1} \frac{R^{(0)}(z)
                 - z R^{(0)}{}' (z)}{\[z + R^{(0)}(z) + w\]^p}           \\
                \~\Phi_n^p(w)
            &=& \frac{1}{\Gamma(n)} \int_0^\infty\! dz\, z^{n-1} \frac{R^{(0)}(z)
                                   }{\[z + R^{(0)}(z) + w\]^p}
                \end{eqnarray*}}\hfill
\parbox{7mm}{\equa{4.12}{}}\newline
for $n>0$. It follows that $\Phi_0^p(w) = \~\Phi_0^p(w) = \(1+w\)^{-p}$ for
$n=0$. In addition we use the fact that
\equ{4.14}{{\cal N} = \frac{\partial_t\!\[ \Z\, k^2 R^{(0)}(- D^2/k^2)
            \]}{\Z} = \[2 - \eta_N (k)\] k^2 R^{(0)} + 2 D^2 R^{(0)}{}'}
with $\eta_N(k) = -\partial_t (\ln \Z) = - \partial_t (\Z)/\Z$ being the
anomalous dimension of the operator $\sqrt{g} R$. Then we can rewrite the
equations (\ref{e4.10}) and (\ref{e4.11}) in terms of $\Phi$ and $\~\Phi$. This
leads to the following system of equations
\equ{4.15}{\begin{array}{r@{}l}
            \D\partial_t\!\(\Z \lamb\) = \frac{1}{4\kappa^2}\,\frac{k^4}{\(4\pi\)^2}
            \bigg\{  & 6\,  \Phi_2^1\!\(-2        \lamb/k^2\) +
                       4\,  \Phi_2^1\!\(-2 \alpha \lamb/k^2\) -
                       8\,  \Phi_2^1\!\(                  0\)         \\[1.1em]
  \D-\eta_N(k)\Big[  & 3\,\~\Phi_2^1\!\(-2        \lamb/k^2\) +
                       2\,\~\Phi_2^1\!\(-2 \alpha \lamb/k^2\) \Big] \bigg\} \quad,
           \end{array}}
\equ{4.16}{\begin{array}{r@{}l}
           \D\partial_t \Z = - \frac{1}{12\kappa^2}\,\frac{k^2}{\(4\pi\)^2}
         \bigg\{&-\,   30\,            \Phi_2^2\!\(-2        \lamb/k^2\) -
                     6(2\alpha-1)\,    \Phi_2^2\!\(-2 \alpha \lamb/k^2\) \\[1.1em]
                &+\,\:\;6\,            \Phi_1^1\!\(-2        \lamb/k^2\) +
                        4\,            \Phi_1^1\!\(-2 \alpha \lamb/k^2\) \\[1.1em]
                &-\,   12\,            \Phi_2^2\!\(                  0\) -
                        8\,            \Phi_1^1\!\(                  0\) \\[1.1em]
\D-\eta_N(k)\Big[&-\,  15\,          \~\Phi_2^2\!\(-2        \lamb/k^2\) -
                     3(2\alpha-1)\,  \~\Phi_2^2\!\(-2 \alpha \lamb/k^2\) \\[1.1em]
                &+\,\:\;3\,          \~\Phi_1^1\!\(-2        \lamb/k^2\) +
                        2\,          \~\Phi_1^1\!\(-2 \alpha \lamb/k^2\)
           \Big] \bigg\} \quad.
           \end{array}}

Now we introduce the dimensionless, renormalised Newtonian and cosmological 
constants
\equ{4.17}{g_k = k^2\,G_k = k^2\,\Z^{-1}\,\Gb, \quad \lam = k^{-2}\,\lamb \quad.}
Here $G_k$ is the renormalised but dimensionful Newtonian constant at scale $k$.
The evolution equation for $g_k$ reads then
\equ{4.18}{\partial_t g_k = \[2+\eta_N(k)\] g_k \quad.}
From (\ref{e4.16}) we find the anomalous dimension $\eta_N(k)$
\equ{4.19}{\eta_N(k) = g_k B_1(\lam) + \eta_N(k) g_k B_2(\lam)}
where we used the notation
\newline
\parbox{13.9cm}{\begin{eqnarray*}
                B_1(\lam)
 &=&                \frac{1}{3 \pi} \[
                   3\,  \Phi_1^1(-2        \lam) +
                   2\,  \Phi_1^1(-2 \alpha \lam) -
                   4\,  \Phi_1^1( 0            )     \right.   \\
 & & \left.\;\: - 15\,  \Phi_2^2(-2        \lam) -
       3\(2\alpha-1\)   \Phi_2^2(-2 \alpha \lam) -
                   6\,  \Phi_2^2( 0            ) \]            \\
                B_2(\lam)
 &=&               -\frac{1}{6 \pi} \[
                   3\,\~\Phi_1^1(-2        \lam) +
                   2\,\~\Phi_1^1(-2 \alpha \lam)     \right.   \\
&&\left.\quad\,\,-15\,\~\Phi_2^2(-2        \lam) -
       3\(2\alpha-1\) \~\Phi_2^2(-2 \alpha \lam) \]  \quad.
                \end{eqnarray*}}\hfill
\parbox{7mm}{\equa{4.20}{}}\newline
Solving (\ref{e4.19})
\equ{4.21}{\eta_N(k) = \frac{g_k B_1(\lam)}{1- g_k B_2(\lam)} }
we see that the anomalous dimension $\eta_N$ is a nonperturbative quantity.
From (\ref{e4.15}) we obtain the evolution equation for the cosmological
constant
\equ{4.22}{\begin{array}{r@{}l}
           \D\partial_t \lam = -\[2-\eta_N(k)\] \lam + \frac{1}{2\pi}\, g_k
           \Big\{&\D 6\,  \Phi_2^1(-2        \lam) +
                     4\,  \Phi_2^1(-2 \alpha \lam) -
                     8\,  \Phi_2^1( 0            )               \\[1.1em]
-\eta_N(k) \Big[ &\D 3\,\~\Phi_2^1(-2        \lam) +
                     2\,\~\Phi_2^1(-2 \alpha \lam)   \Big] \Big\}
           \end{array}}
The equations (\ref{e4.18}) and (\ref{e4.22}) together with (\ref{e4.21}) give 
the system of differential equations for the two $k$-depending constants $\lam$
and $g_k$. If the initial values of $g_\Lambda$ and $\lambda_\Lambda$ are known
it determines the running of these constants for $k \leq \Lambda$. This system
contains effects of arbitrarily high orders in the perturbation theory.
In the special case $\alpha = 1$ we get back the result of Reuter in \cite{a}.

\ssection{The Running of the Newtonian constant}
We will calculate now the $\alpha$-dependence of the running Newtonian constant.
To do this we start from the evolution equation for the renormalised
gravitational constant
\equ{5.1}{\partial_t G_k = \eta_N(k)\, G_k }
with $\eta_N$ taken from (\ref{e4.21}) and $g_k = k^2 G_k = k^2 \Gb + O(\Gb^2)$.
Let us assume that the cosmological constant is much smaller than the IR cutoff
scale, $\lamb \ll k^2$. So we can neglect all contribution of $\lam$ in
$\eta_N$ to simplify the result. Then the evolution equation has the
solution
\equ{5.6}{G_k = G_0 \[1 - \sigma\, \Gb\, k^2 \]^{\omega/\sigma} }
with
\equ{5.7}{\sigma =  B_2(0) = \frac{3 \alpha + 1}{6 \pi}, \quad
          \omega = -\frac{B_1(0)}{2}
                 =  \frac{\pi}{36}\[\frac{36(\alpha + 3)}{\pi^2} - 1\]\quad.}
We see that $\omega > 0$ for $\alpha > \frac{\pi^2}{36}-3 \approx -2.73$. This
means the gravitation is "anti\-screening" in such a way that $G_k$ decreases 
if $k$ increases, i.~e. with increasing distance also the Newtonian constant 
increases. This supports the naive picture that the mass of a particle 
receives additional positive contributions from the virtual particles 
surrounding it.

In addition there is a second interval for $\alpha$ where the antiscreening
changes to a screening effect. The coefficient $\sigma$ does not play any role
in the change from antiscreening to screening. The significant influence of
the running of $G_k$ comes from the sign of the numerator of the exponent in
(\ref{e5.6}) which is $\omega$. The value $\alpha = \frac{\pi^2}{36}-3$ is
significant for the used truncation and zeroth order in $\lam$.

This screening behaviour is an artifact of the truncation where we used
$\alpha(k)=\alpha=const$. A better truncation would include $\alpha$ as
$k$-depending parameter as it is done in \cite{x} for pure Yang-Mills theory.
There it has been shown that $\alpha=0$ is an attracting fixed point of the
truncated evolution. This is to be expected for gravity too. 
From (\ref{e4.16}) we see that $\Z \sim k^2$. As in Yang-Mills theory we 
expect that $\alpha/\Z$ is a  constant and does not depend on $k$. Then also 
$\alpha$ should tend to zero quadratically for $k \to 0$.

To get back the result in \cite{a} we expand (\ref{e5.6}) with respect to
$\Gb k^2$ and obtain
\equ{5.3}{G_k = G_0 \[1 - \omega\, \Gb\, k^2 + O\!\(\Gb^2 k^4\)\] }
with $\omega$ given by (\ref{e5.7}). For $\alpha=1$ Reuters result is obtained.

\begin{figure}[tb]
\centerline{\epsfig{figure=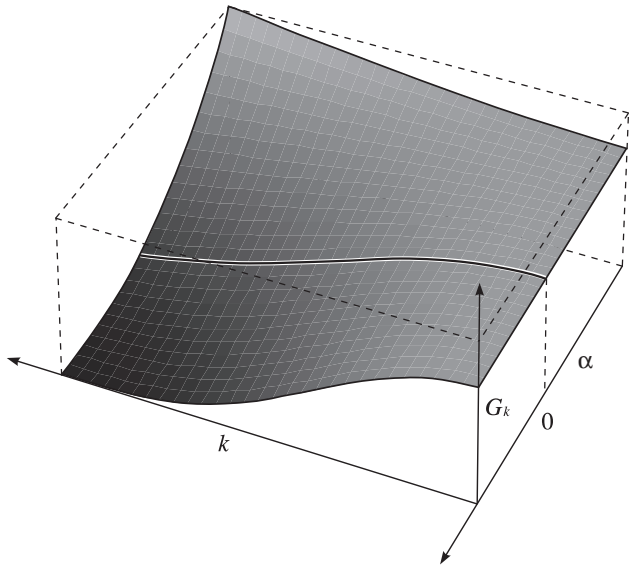, scale=.75}}
\caption{The gravitational constant as function of $k$ and $\alpha$.}
\label{f1}
\end{figure}
In figure \ref{f1} the gravitational constant $G_k$ is shown as function of
the gauge parameter $\alpha$ and the cutoff scale $k$. The Landau-De Witt gauge
$\alpha = 0$ is drawn as a separate curve on the surface. Both regions ---
antiscreening and screening region --- are to perceive.

\ssection{Discussion}
In order to solve the gauge dependence problem we suggest two ways. The first
way is the use of a better truncation which includes the gauge parameter as a 
function of $k$. As argued in the previous section $\alpha=0$ should be a 
stable fixed point of the theory. It would be also interesting to use more 
general gauge fixing conditions.

Alternatively it is possible to work with a generalisation of the standard
effective action, the so-called gauge-fixing independent effective action
(for an introduction and its different forms, see \cite{s, c}). The gauge-fixing
independent effective action leads to the same $S$-matrix as the standard one;
but by construction it does not depend on the choice of the gauge condition
(off-shell it differs from the standard effective action). Moreover, it has
been shown \cite{t} that in Einstein gravity on De Sitter background the
one-loop standard effective action in the Landau-De Witt gauge $\alpha=0$
coincides with the one-loop gauge-fixing independent effective action \cite{s}.

Working with the truncated evolution equation which is very similar to the
one-loop approach one may give arguments that the truncated scale-dependent
gravitational effective action at $\alpha=0$ coincides with the gauge-fixing
independent one\footnote{Note that the latter effective action is not
constructed yet as complete scheme.}. Then the physical value of $\omega$
would be given by (\ref{e5.7}) at $\alpha=0$ again.

As one sees the Einstein gravity looks as an antiscreening theory where the
gravitational constant increases at large distances:
\equ{6.1}{G(r) = G(\infty) \[1 - \omega\, \frac{\Gb \hbar}{r^2 c^3} +
          O\!\(\frac{\Gb^2}{r^4}\) \]  }
where $r=k^{-1}$, ($\Gb/r^2 \ll 1$), and the correspondence with the Newtonian
potential may be restored.

Of course, the behaviour of the gravitational coupling constant depends very
much on the type of the model one considers. For example it has a
qualitatively different form for the effective theory of the conformal factor
\cite{j, k}, for $R^2$-gravity \cite{l} or for the four-dimensional dilaton
gravity \cite{j}. In the very recent calculation of $\omega$ in \cite{w} (see
also references therein) it was also found that $\omega$ is positive. Similarly,
in $\alpha=1$ gauge \cite{l} $\omega$ is found to be positive again.

Hence, our study of gauge dependence of truncated evolution equation in
Einstein gravity indicates that the gravitational coupling constant increases 
at large distances.

There are different possibilities to extend the results of the present work.
The next problem is the calculation using an improved truncation. One can also
consider a more complicated theory like gauged supergravity on gravitational
background, to formulate there evolution equation for scale-dependent effective
action, and to study the running of gravitational and cosmological constants. We
again will find the problem of the gauge dependence of the evolution equation.
It could be also of interest to discuss in details the cosmological applications
of gravitational and cosmological constants running (see \cite{t}).

Finally, one has to note that taking into account of combined gravity-matter
effects (unified in supermultiplet as in supergravities theories or just
Einstein-matter type systems) may also change the behaviour of gravitational and
cosmological couplings.

\section*{Acknowledgments}
S.~D.~O.~thanks R.~Percacci and R.~Floreanini for useful discussions and
the Saxonian Min.~of Science (Germany) and COLCIENCIAS (Colombia) for 
partial support of this work.\\
S.~F.~thanks B.~Geyer and M.~Reuter for helpful discussions. This work has 
been supported by DFG, project no.~GRK 52.


\newpage
\begin{thebibliography}{99}
\bibitem{c}I.~L.~Buchbinder, S.~D.~Odintsov and I.~L.~Shapiro, \emph{Effective
           Action in Quantum Gravity}, IOP, Bristol (1992).
\bibitem{h}I.~Antoniadis and E.~Mottola, \emph{Phys.~Rev.~D} \textbf{45}, 2013 (1992).\\
           S.~D.~Odintsov, \emph{Z.~Phys.~C} \textbf{54}, 531 (1992).
\bibitem{i}N.~C.~Tsamis and R.~P.~Woodard, \emph{Ann.~Phys. (NY)} \textbf{238}, 1 (1995).\\
           A.~D.~Dolgov, M.~B.~Einhorn and V.~I.~Zakharov, \emph{Phys.~Rev.~D} \textbf{52}, 
           717 (1995).
\bibitem{j}E.~Elizalde, A.~G.~Jacksenaev, S.~D.~Odintsov and I.~L.~Shapiro,
           \emph{Class.~Quant. Grav.} \textbf{12}, 1385 (1995).
\bibitem{m}E.~Elizalde and S.~D.~Odintsov, \emph{Phys.~Lett.~B} \textbf{315}, 245 (1993).
\bibitem{n}I.~L.~Buchbinder and A.~Y.~Petrov, \textbf{hep-th/9607093}, to appear in
           \emph{Class. Quant.~Grav.} (1996).
\bibitem{k}I.~Antoniadis and S.~D.~Odintsov, \emph{Phys.~Lett.~B} \textbf{343}, 76 (1995).\\
           E.~Elizalde, S.~D.~Odintsov and I.~L.~Shapiro, \emph{Class.~Quant.~Grav.}
           \textbf{11}, 1607 (1994).
\bibitem{l}E.~Elizalde, C.~Lousto, S.~D.~Odintsov and A.~Romeo, \emph{Phys.~Rev.~D}
           \textbf{52}, 2202 (1995).
\bibitem{o}K.~G.~Wilson and J.~Kogut, \emph{Phys.~Rep.} \textbf{12}, 75 (1974).\\
           F.~Wegner and A.~Houghton, \emph{Phys.~Rev.~A} \textbf{8}, 401 (1973).\\
           J.~Polchinski, \emph{Nucl.~Phys.~B} \textbf{231}, 269 (1984).
\bibitem{d}C.~Wetterich, \emph{Phys.~Lett.~B} \textbf{301}, 90--94 (1993).
\bibitem{e}M.~Reuter and C.~Wetterich, \emph{Nucl.~Phys.~B} \textbf{417}, 181--214 (1994).
\bibitem{f}M.~Reuter and C.~Wetterich, \emph{Nucl.~Phys.~B} \textbf{427}, 291--324 (1994).
\bibitem{p}M.~Reuter and C.~Wetterich, \emph{Nucl.~Phys.~B} \textbf{391}, 147 (1993).\\
           M.~Reuter, \emph{Phys.~Rev.~D} \textbf{53}, 4430 (1996).
\bibitem{q}M.~Bonini, M.~D'Attanasio and G.~Marchesini, \emph{Nucl.~Phys.~B}
           \textbf{418}, 81 (1994).\\
           M.~Bonini, M.~D'Attanasio and G.~Marchesini, \emph{Nucl.~Phys.~$\!$B}
           \textbf{437}, 163 (1995).\\
           U.~Ellwanger and L.~Vergara, \emph{Nucl.~Phys.~B} \textbf{398}, 52 (1993).\\
           M.~Alford and J.~March-Russel, \emph{Nucl.~Phys.~B} \textbf{417}, 527 (1994).\\
           S.~B.~Liao, J.~Polonyi and D.~Xu, \emph{Phys.~Rev.~D} \textbf{51}, 748 (1995).\\
           T.~R.~Morris, \emph{Phys.~Lett.~B} \textbf{329}, 241 (1994).
\bibitem{a}M.~Reuter, \tb{hep-th/9605030} (1996).
\bibitem{r}R.~Floreanini and R.~Percacci, \emph{Nucl.~Phys.~B} \textbf{436}, 141 (1995).\\
           R.~Floreanini and R.~Percacci, \emph{Phys.~Rev.~D} \textbf{52}, 896 (1995).
\bibitem{u}F.~Freire and C.~Wetterich, \emph{Phys.~Lett.~B} \textbf{380}, 337 (1996).
\bibitem{x}U.~Ellwanger, M.~Hirsch and A.~Weber, \emph{Z.~Phys.~C} \textbf{69}, 687 (1996).
\bibitem{b}E.~S.~Fradkin and A.~A.~Tseytlin, \emph{Nucl.~Phys.~B} \textbf{234}, 472--508 (1984).
\bibitem{g}G.~W.~Gibbons and M.~J.~Perry, \emph{Nucl.~Phys.~B} \textbf{146}, 90--108 (1978).\\
           S.~M.~Christensen and M.~J.~Duff, \emph{Nucl.~Phys.~B} \textbf{170}, 480--506 (1980).
\bibitem{s}G.~A.~Vilkovisky, \emph{Nucl.~Phys.~B} \textbf{234}, 125 (1984).\\
           B.~S.~De~Witt in: \emph{Architecture of Fundamental Interactions at
           Short Distances}, Eds.: S.~Ramond and R.~Stora, North-Holland Publ.~Co. (1987).\\
           C.~P.~Burgess and G.~Kunstatter, \emph{Mod.~Phys.~Lett.~A} \textbf{2}, 875 (1987).\\
           P.~M.~Lavrov, S.~D.~Odintsov and I.~V.~Tyutin, \emph{Mod.~Phys.~Lett.~A}
           \textbf{3}, 1273 (1988).\\
           S.~D.~Odintsov, \emph{Fortschr.~Phys.} \textbf{38}, 371 (1990).\\
           G.~Kunstatter, \emph{Class.~Quant.~Grav.} \textbf{9}, 157 (1992).
\bibitem{t}E.~S.~Fradkin and A.~A.~Tseytlin, \emph{Nucl.~Phys.~B} \textbf{234}, 509 (1984).\\
           I.~L.~Buchbinder, E.~N.~Kirillova and S.~D.~Odintsov, \emph{Mod.~Phys.~Lett.~A}
           \textbf{4}, 633 (1989).\\
           S.~D.~Odintsov, \emph{Europhys.~Lett.} \textbf{10}, 287 (1989).\\
           S.~D.~Odintsov, \emph{Th.~Math.~Phys.~USSR} \textbf{82}, 66 (1990).\\
           T.~R.~Taylor and G.~Veneziano, \emph{Nucl.~Phys.~B} \textbf{345}, 210 (1990).\\
           H.~T.~Cho and R.~Kantowski, \emph{Phys.~Rev.~Lett.} \textbf{67}, 422 (1991).
\bibitem{w}H.~W.~Hamber and S.~Liu, \emph{Phys.~Lett.~B} \textbf{357}, 51 (1995).
\end{thebibliography}
\end{document}